\documentclass[12pt]{article}
\usepackage{mathrsfs}
\usepackage{natbib,graphicx,setspace,lscape,longtable,epstopdf,xr}
\usepackage{mathrsfs,amsmath,amsthm,amssymb,color}
\usepackage{natbib,epsfig,graphicx,pdfpages}
\usepackage{rotating}
\usepackage{float}
\usepackage{bm}
\usepackage{ulem}
\usepackage{CJK}
\usepackage{ragged2e}
\usepackage{threeparttable}
\usepackage{multirow}
\usepackage{url}
\newcommand{\tabincell}[2]{\begin{tabular}{@{}#1@{}}#2\end{tabular}}

\bibpunct{(}{)}{;}{a}{,}{,}

\newtheorem{theorem}{Theorem}
\newtheorem{lemma}{Lemma}
\newtheorem{proposition}{Proposition}

\newcommand{\csection}[1]
    {\begin{center}
        \stepcounter{section}
        {\bf\large\arabic{section}. #1}
    \end{center}
}

\newcommand{\scsection}[1]
    {\begin{center}
        {\bf\large #1}
    \end{center}
}

\newcommand{\csubsection}[1]{
\begin{center}
\stepcounter{subsection}
{\it\arabic{section}.\arabic{subsection}. #1}
\end{center}
}

\newcommand{\scsubsection}[1]{
\begin{center}
\stepcounter{subsection}
{\it #1}
\end{center}
}

\def\n{\nonumber}

\def\beq{\begin{equation}}
\def\eeq{\end{equation}}
\def\beqr{\begin{eqnarray}}
\def\eeqr{\end{eqnarray}}
\def\beqrs{\begin{eqnarray*}}
\def\eeqrs{\end{eqnarray*}}
\def\bet{\begin{theorem}}
\def\eet{\end{theorem}}
\def\bel{\begin{lemma}}
\def\eel{\end{lemma}}
\def\bep{\begin{proposition}}
\def\eep{\end{proposition}}
\def\bg{\begin{figure}[tbph]\begin{center}}
\def\eg{\end{center}\end{figure}}

\def\bc{\begin{center}}
\def\ec{\end{center}}

\def\bP{\mathcal P}

\def\mR{\mathbb{R}}

\def\mS{\mathbb S}

\def\pM{\mathbf M}
\def\mS{\mathcal S}

\def\bE{\mathbb E}

\def\var{\mbox{var}}

\renewcommand{\arraystretch}{1.3}

\numberwithin{equation}{section}

\begin{document}
\begin{center}
{\bf\Large Efficient Estimation for Generalized Linear Models on a Distributed System with Nonrandomly Distributed Data}\\
\bigskip

Feifei Wang$^{1,2}$, Danyang Huang$^{1,2}$, Yingqiu Zhu$^1$, and Hansheng Wang$^3$

{\it\small
$^1$ Center for Applied Statistics, Renmin University of China, Beijing, China;\\
$^2$ School of Statistics, Renmin University of China, Beijing, China;\\
$^3$ Guanghua School of Management, Peking University, Beijing, China.
}

\end{center}

\begin{footnotetext}[1]
{
Feifei Wang's research is supported by building world-class universities (disciplines) of
the Renmin University of China, the Fundamental Research Funds for the
Central Universities and the Research Funds of Renmin University of China
(No. 18XNLG02);
Danyang Huang is supported by National Natural Science Foundation of China
(NSFC, 11701560), building world-class universities (disciplines) of
the Renmin University of China, the Fundamental Research Funds for the
Central Universities and the Research Funds of Renmin University of China
(No. 18XNLG02);
Hansheng Wang is partially supported by National Natural Science Foundation of China (No. 71532001, No. 11525101, No. 71332006).  It is also supported in part by China's National Key Research Special Program (No. 2016YFC0207700).
}
\end{footnotetext}

\begin{singlespace}
\begin{abstract}

Distributed systems have been widely used in practice to accomplish data analysis tasks of huge scales. In this work, we target on the estimation problem of generalized linear models on a distributed system with nonrandomly distributed data. We develop a Pseudo-Newton-Raphson algorithm for efficient estimation. In this algorithm, we first obtain a pilot estimator based on a small random sample collected from different Workers. Then conduct one-step updating based on the computed derivatives of log-likelihood functions in each Worker at the pilot estimator. The final one-step estimator is proved to be statistically efficient as the global estimator, even with nonrandomly distributed data. In addition, it is computationally efficient, in terms of both communication cost and storage usage. Based on the one-step estimator, we also develop a likelihood ratio test for hypothesis testing. The theoretical properties of the one-step estimator and the corresponding likelihood ratio test are investigated. The finite sample performances are assessed through simulations. Finally, an American Airline dataset is analyzed on a Spark cluster for illustration purpose.

\end{abstract}


\end{singlespace}

\newpage

\csection{Introduction}

With the rapid development of technology, massive data are often encountered in both scientific fields and daily life \citep{gopal2013distributed,battey2015distributed}.
In many cases, such a huge dataset is usually hard to be efficiently dealt by one single computer, but needs a large set of connected computers, which are referred to as a distributed system \citep{duchi2014optimality}. Here, we consider a standard ``master-and-worker" architecture for distributed systems, which is the most popularly used type in practice (e.g., Hadoop, Spark, etc). Under this architecture, a computer serves as the Master, while all the other computers serve as Workers. The analysis task of a huge dataset is then divided into pieces, each of which is solved by one Worker. Therefore, the use of distributed systems enable us to accomplish data analysis
tasks of huge scales \citep{andrews2000foundations,gopal2013distributed,battey2015distributed}.

In distributed systems, two problems are often worth of consideration. The first one is \textit{communication cost}, which refers to the wall-clock time from inter-computer communication. In a standard ``master-and-worker" distributed system, inter-computer communication is only allowed between the Master and the Workers. Therefore, the communication cost among different Workers is often expensive \citep{duchi2014optimality,fan2017distributed,jordan2018communication}. The second problem which needs consideration is the storage limitation. For example, the popular distributed system Apache Spark uses \textit{resilient distributed dataset (RDD)} as its foundation, which requires data and intermediate results to be cached in memory during computation \citep{zaharia2010spark,zaharia2012resilient,gu2013memory}. This feature enables Spark to speed up the computational time, but have to pay for the high memory cost of RDDs \citep{gu2013memory,chand2017analysis}. Therefore, how to improve the memory usage in distributed computing is also a problem of interest.



Except for the communication cost and memory usage, another important issue is the distribution of data on a distributed system. In an ideal situation, the massive data should be randomly distributed on different Workers \citep{zhang2012communication}. Consequently, any statistics computed from a single Worker should be a consistent estimate for its global counterpart. However, this is seldom the case in practice. Actually, practitioners usually store a huge dataset in a way convenient for operation. For example, the data might be distributed on different Workers by location or time. As a consequence, the observed distribution of data is different across Workers, and thus the resulting statistics calculated on different Workers could be seriously biased for its global counterpart. If this is the case, the final estimate, summarizing from those computed on different Workers, would be questionable.

In this work, we target on the estimation problem of generalized linear models (GLM) on a distributed system, and try to achieve the following goals: reduce the communication cost, increase the memory efficiency, and consider the nonrandomly distributed nature of data. The generalized linear model is a broad class of models that generalize the linear regression to allow for various types of response and have wide applications \citep{mccullagh1989generalized,dobson2008introduction,fahrmeir2013multivariate}. To estimate GLM on a distributed system, the commonly used Newton-Raphson algorithm cannot be applied directly, due to the communication and memory cost in multiple iterations.

To address this problem, researchers have developed a number of effective solutions. One popular method is the so-called one-shot estimate (or mixture average, \citealt{mcdonald2009efficient,zinkevich2010parallelized,zhang2012communication}). The idea of this method is to conduct the estimation of  GLM on each Worker to obtain a local estimate, and the final global estimate computed by the Master is a simple average of the local ones. It is communication-efficient, but may lead to inconsistent global estimate when the data is nonrandomly distributed in different Workers. In addition, it could not obtain the best statistical estimation efficiency in many cases \citep{Shamir:2014,jordan2018communication}. To improve the estimation efficiency, iterative algorithms, which need multiple rounds of communications, have been proposed \citep{boyd2011distributed,Shamir:2014,wang2017efficient,jordan2018communication}.





All these methods have been proven practically useful, but under one critical assumption, i.e., the data is randomly distributed on different Workers.
However, as we mentioned before, this is seldom the case in practice. In fact, practitioners often find the local estimates received from different Workers are quite different from each other. This can be viewed as a strong evidence that the data distributed on different Workers might be quite different. As a consequence, the local estimates may not be consistent, thus leads to questionable global estimate in one-shot method.
Then, how to obtain an efficient estimate for GLM on a distributed system with nonrandomly distributed data becomes a problem of interest.


To address this issue, we develop here a Pseudo-Newton-Raphson algorithm for an efficient estimation of generalized linear models on a distributed system with nonrandomly distributed data. Specifically, in the first step, we collect a random sample of small size $n$ from different Workers in the distributed system. The MLE $\tilde\theta$ is calculated as a pilot estimate based on this sample, which is $\sqrt{n}$-consistent and requires small memory usage.
In the second step, we broadcast the pilot estimate $\tilde\theta$ to each Worker, and then calculate the first and second order derivatives of the log-likelihood function in each Worker at the point $\tilde\theta$. Then, the computed derivatives in each Worker is received by the Master with little communication cost, since they are finite dimensional vectors or matrices. Finally, the Master combines the received derivatives to compute a global estimate, and conducts one-step updating (e.g., the Newton-Raphson type) with $\tilde\theta$ as the initial estimate. This leads to the final estimate $\hat\theta$.

It is noteworthy that, the above updating method only requires two rounds of Master-and-Worker communication, and the transferred data are all of small sizes. Therefore, this method is \textit{computationally efficient}, in terms of both communication cost and storage usage. In addition, we prove that the final estimate $\hat\theta$ shares the same asymptotic covariance as the global MLE as long as $n^2\gg N$, which suggests it is also \textit{statistically efficient}. Based on the final estimate, we also develop a likelihood ratio test for hypothesis testing on a distributed system with nonrandomly distributed data.

The rest of this paper is organized as follows. Section 2 introduces the efficient estimation for GLM and the corresponding likelihood ratio test. Section 3 presents a number of simulation studies to demonstrate the finite sample performances of our proposed estimator and the likelihood ratio test. Section 4 illustrates the application of our method using the Airline dataset (\url{http://stat-computing.org/dataexpo/2009/}). Section 5 concludes the paper with a brief discussion.

\csection{Efficient Estimation for Generalized Linear Models}
\csubsection{Maximum Likelihood Estimation}

Consider a sample with $N$ observations. For each observation $i$ $(1\leq i \leq N)$, we collect a response $Y_i$ and a $d$-dimensional covariate vector $X_i=(X_{i1},\cdots,X_{id})\in\mR^d$. We assume that $(X_i^\top,Y_i)$s are independent and identically distributed. Suppose that the sample size $N$ is large and the data are stored on a distributed system with $K$ local Workers. Define $\mS=\{1,\cdots, N\}$, and further define  $\mS_k$ to be the set of sample indices stored on the $k$-th $(1\leq k \leq K)$ Worker with $|\mS_k|=N_k$. Then we have $\mS_{k_1}\cap \mS_{k_2}=\emptyset$ for $k_1\neq k_2$, $\mS=\cup_{k=1}^K\mS_k$, and $N=\sum_{k}N_k$.

We define the generalized linear models according to the previous literature \citep{Nelder:1972,Wedderburn:1974,McCullagh:2019}. Define the expectation of $Y_i$ to be $\mu_i$. And $\theta_i$ is referred to as the {\it canonical parameter}, which is some function of $\mu_i$. We assume that the distribution of $Y_i$ is in an exponential family, whose log-likelihood could be spelled out as,
\beqrs
\ell(Y_i,\theta_i,\phi)=\big\{Y_i\theta_i-g(\theta_i)\big\}/a(\phi)+c(Y_i,\phi),
\eeqrs
where $\phi$ is a nuisance parameter, and $a(\cdot)$ and $c(\cdot)$ are some specific functions.
Define $\eta_i=\sum_{j=1}^d\beta_jX_{ij}$ to be the linear predictor based on the covariate vector, where $\beta=(\beta_1\cdots\beta_d)\in\mR^d$ is a $d$-dimensional unknown parameter. If we define $\theta_i=\eta_i$, then a {\it canonical link} relates the linear predictor $\eta_i$ to the expected value $\mu_i$. For example, for logistic regression, the canonical link could be $\eta=\log\{\mu/(1-\mu)\}$; for poisson regression, the canonical link is $\eta=\log(\mu)$.

The unknown parameter $\beta$ is often estimated by the maximum likelihood estimator (MLE), which is denoted as $\hat\beta_{M}$. It could be obtained by maximizing the log-likelihood function $\ell_f(\beta)$, namely,
\beqrs
\hat\beta_M=\arg\max_{\beta}\ell_f(\beta)=\arg\max_{\beta}\sum_{i=1}^N \big\{Y_i\theta_i-g(\theta_i)\big\}/a(\phi),
\eeqrs
where some constants are ignored.
Since there is no general closed-form solution to MLE, it could be obtained by Newton's method.

\noindent
\textbf{Remark 1.} Specifically, define $V(\cdot)$ is a function which satisfies $\var(Y_i)=a(\phi)V(\mu_i)$, so that $V(\mu_i)$ is the variance of $Y_i$ when the scale factor $a(\phi)$ is unity. For the sake of notation simplification, we assume that $a(\phi)=1$ throughout the article. The results could be similarly generalized for the cases where $a(\phi)\neq 1$.

Next we consider the theoretical properties of the estimator $\hat\beta_M$. Under certain conditions, it has been shown that $\hat\beta_M$ is $\sqrt{N}$-consistent and asymptotically normal \citep{McCullagh:1989}.  We further define $\pM=E\big[(\partial \mu_i/\partial \beta)\big\{V(\mu_i)\big\}^{-1}(\partial \mu_i/\partial \beta)^\top\big]$, which is a positive definite matrix, then we have,
\beqr
\sqrt{N}(\hat\beta_M-\beta_0)\stackrel{d}{\longrightarrow}N(\mathbf{0}_d,\pM^{-1}).
\eeqr
where $\mathbf{0}_d\in\mR^d$ is a zero vector, and $\beta_0$ is the true parameter. See \cite{Nelder:1972,Wedderburn:1974,McCullagh:2019} for more technical details.


\csubsection{One-Step Estimation}

The one-step estimation consists of two steps. In the first step, a sample of small size $n$ is randomly selected from different Workers as a {\it pilot sample}. It is assumed that $n$ is much smaller than $N$, which is $n/N\rightarrow 0$ but $n\rightarrow \infty$ as $N\rightarrow \infty$. Specifically, let $\bP_k$ be the indices selected from $\mS_k$ in the $k$-th Worker by simple random sampling without replacement, and $|\bP_k|=n_k$. Thus the pilot sample could be denoted as $\bP=\cup_k \bP_k$. Since the pilot sample size $n$ is much smaller than $N$, the communication cost in transferring the pilot sample from different Workers to the Master is practically acceptable. Then a {\it pilot estimator} could be obtained on the Master based on the pilot sample, which is denoted as,
\beqr
\hat\beta_p=\arg\max_\beta\ell_p(\beta)=\sum_{i\in \bP}\big\{Y_i\theta_i-g(\theta_i)\big\},
\eeqr
where $\ell_p(\beta)$ is the log-likelihood function based on the pilot sample. It is remarkable that $\theta_i$ is a function of $\beta$. Specifically, the pilot sample is obtained in a completely random manner. Thus $\hat\beta_p$ is consistent regardless of how the data are distributed on different Workers. However, $\hat\beta_p$ is statistically inefficient because it is $\sqrt n$-consistent.

Recall that $\hat\beta_M$ is obtained based on the whole sample. It could be expressed that,
\beqr
\hat\beta_M=\beta_0+\Big\{\frac{1}{N}\sum_{i=1}^N\frac{1}{V(\mu_{i0})}\dot\mu_{i}\dot\mu_{i}^\top\Big\}^{-1} \Big\{\frac{1}{N}\sum_{i=1}^N\frac{(Y_i-\mu_{i0})}{V(\mu_{i0})}\dot\mu_{i}\Big\}+o_p(\frac{1}{\sqrt{N}}),\label{eq:onestep}
\eeqr
where $\mu_{i0}=\mu_{i}|_{\beta=\beta_0}$, $\dot\mu_{i}\equiv(\partial \mu_i/\partial \beta)|_{\beta=\beta_0}$. See Appendix A for more details.
Then, to further improve the efficiency of the estimator, we replace $\beta_0$ in (\ref{eq:onestep}) with $\hat\beta_p$ to obtain $\hat\mu_{pi}$, i.e., $\dot\mu_{pi}=(\partial \mu_i/\partial \beta)|_{\beta=\hat\beta_p}$. This leads to the one-step estimator as $\hat\beta_o$, which is
\beqr
\hat\beta_o=\hat\beta_p+\Big\{\frac{1}{N}\sum_{i=1}^N\frac{1}{V(\hat\mu_{pi})}\dot\mu_{pi}\dot\mu_{pi}^\top\Big\}^{-1} \Big\{\frac{1}{N}\sum_{i=1}^N\frac{(Y_i-\hat\mu_{pi})}{V(\hat\mu_{pi})}\dot\mu_{pi}\Big\}.\label{eq:onestep1}
\eeqr
This one-step estimator $\hat\beta_o$ is in the similar spirit as classical one-step estimator \citep{Shao:2003,Zou:Li:2008}. However, the key difference is that the pilot estimator $\hat\beta_p$ in our one-step estimator is $\sqrt{n}$-consistent. As a consequence, the asymptotic properties of the final estimator $\hat\beta_o$ are more difficult to be studied than the traditional one-step estimator. Moreover, we are able to prove that the one-step estimator has the same asymptotic covariance as the MLE $\hat\beta_M$ under certain conditions.

To establish the statistical properties of the one-step estimator, the following conditions are needed.
\begin{itemize}
\item [(C1)] The pilot sample size $n$ satisfies that $n/N\rightarrow 0$ and $n^2/N\rightarrow \infty$.
\item [(C2)] Assume $E||X_i||<\infty$, and $\pM=E\big[(\partial \mu_i/\partial \beta)\big\{V(\mu_i)\big\}^{-1}(\partial \mu_i/\partial \beta)^\top\big]$ to be a positive definite matrix.
\item [(C3)] Assume there exists a function $M(\cdot)$, such that for any $\beta \in \Theta$, $EM(Y_i)<K$, where $K$ is a constant. And assume that in the neighbourhood of the true parameter $\beta_0$, we have for $j_1,j_2,j_3 \in \{1,\cdots,d\}$,
\beqrs
\left|\frac{\partial\ell (Y_i,\beta)}{\partial \beta_{j_1}\partial \beta_{j_2}\partial \beta_{j_3}}\right|\leq M(Y_i).
\eeqrs
\end{itemize}

\noindent
Condition (C1) considers the relationship between $n$ and $N$. It is assumed that $1/n$ is $o_p(1/\sqrt{N})$. Condition (C2) and (C3) are typical assumptions of the MLE method for GLM. With the condition satisfied, we have the following theorem.
\bet\label{thm1}
Assume the conditions (C1)--(C3) hold, then we have $N^{-1/2}(\hat\beta_o-\beta_0)\stackrel{d}{\longrightarrow}N(\mathbf{0}_d,\pM^{-1}).$
\eet
\noindent
See Appendix B for detailed proof of Theorem \ref{thm1}. From the theorem, we could conclude that $\hat\beta_o$ has the same asymptotic properties as $\hat\beta_M$. Precisely, the difference between $\hat\beta_o$ and $\hat\beta_M$ is $o_p(N^{-1/2})$. However, the computational cost of obtaining global MLE is much higher than that of the one-step estimator, since the former needs multiple iterations, which will lead to inevitable higher communication cost.

\csubsection{Likelihood Ratio Test}

In this section, we consider the hypothesis testing on a distributed system with nonrandomly
distributed data.  The likelihood ratio test is one of the useful tools in hypothesis testing.
Traditionally, suppose we would like to test the null hypothesis,
 \beqrs
 H_0:\beta=\beta_{0}~~~~   \mbox{versus} ~~~~H_1:\beta\neq\beta_{0}.
  \eeqrs
Under the null hypothesis, we could obtain the maximum likelihood estimator $\hat\beta_{M}$ for $\beta$.
Thus,  the likelihood ratio could be defined as,
  \beq\label{eq:likrt}
  \lambda_M(Y)=\frac{\prod_{i=1}^N p(Y_i,\hat\beta_M)}{\prod_{i=1}^N p(Y_i,\hat\beta_{0})},
  \eeq
  where $p(Y_i,\beta)$ is the likelihood function for $Y_i$.
  Then it could be proved that $2\ln \lambda_M(Y)$ asymptotically follows a $\chi^2(d)$ distribution \citep{VanderVaart:2000,McCullagh:2019}.

  However, as we have mentioned above, the maximum likelihood estimator $\hat\beta_M$ is hard to obtain for massive data on a distributed system. Thus under the null hypothesis,  we consider to use the one-step estimator $\hat\beta_o$ as an approximate.
  As a results, a new test statistic based on one-step estimation could be constructed as,
   \beq\label{eq:likrto}
  \lambda_o(Y)=\frac{\prod_{i=1}^N p(Y_i,\hat\beta_o)}{\prod_{i=1}^N p(Y_i,\beta_{0})}.
  \eeq
It is remarkable that, although  the forms of (\ref{eq:likrt}) and (\ref{eq:likrto}) are almost the same, the difference of used estimators will lead to totally different computational cost. Furthermore, we have the following theorem.
\bet\label{thm2}
Under conditions (C1)--(C3), $2\ln \lambda_o(Y)$  asymptotically follows a $\chi^2(d)$ distribution.
\eet
\noindent
See Appendix C for detailed proof of Theorem \ref{thm2}. By the theorem, we could see that $2\ln \lambda_o(Y)$ has the same asymptotic distribution as $2\ln \lambda_M(Y)$. In the next section, we will illustrate both the performance of $2\ln \lambda_o(Y)$ and $2\ln \lambda_M(Y)$.

\noindent
\textbf{Remark.} If we would like to test part of the parameters of $\beta$, the test statistics by one-step method could be similarly established. For example, we assume $\beta_s=(\beta_{l_k})\in \mR^{s}$ is a subvector of $\beta$ with $l_k\in\{1,\cdots,d\}$. We assume that $1\leq l_k\leq s$ and $\beta$ could be written as $\beta=(\beta_s^\top,\beta_c^\top)^\top$.  The null hypothesis is  H$_{0}:\beta_s=\beta_{s0}$ and the alternative one is H$_1:\beta_s\neq\beta_{s0}$. Under the alternative hypothesis, we could estimate $\beta_{c}$ by its one-step estimator $\hat\beta_{co}$. Then define $\hat\beta_{0,o}=(\beta_{s0}^\top,\hat\beta_{co}^\top)^\top$. Thus the test statistics could be constructed as $ \lambda_o(Y)=\{\prod_{i=1}^N p(Y_i,\hat\beta_{\hat\beta_{0,o}})\}^{-1}\{\prod_{i=1}^N p(Y_i,\hat\beta_o)\}.$ It could be proved that $2\ln \lambda_o(Y)$  asymptotically follows a $\chi^2(s)$ distribution.

\csection{Simulation Studies}

In this section, we would first investigate the finite sample performance of the proposed one-step estimator in Section 3.1 to 3.3. Then, we study the performance of the likelihood ratio test using one-step estimator in Section 3.4.

\csubsection{Simulation Design}

To demonstrate the finite sample performance of the one-step estimator, we present a variety of simulation studies. Assume the whole dataset contain $N$ observations, and be stored in $K$ Workers. Here, different sample sizes and number of Workers would be considered, i.e., $N=(10,20,100)\times 10^{3}$ and $K=(2,5,10)$.
We then generate each observation $X_{i}$ and $Y_i$ ($1\leq i \leq N$) under certain generalized linear regression models.
For illustration purpose, we take two classical models as examples, i.e., the logistic regression and poisson regression. The specific settings for these two examples are given as follows.

{\sc Example 1. (Logistic Regression)} The logistic regression deals with analytical tasks with binary responses. In this example, we consider $d = 3$ exogenous covariates $X_{i}=(X_{i1},X_{i2},X_{i3})^{\top}$. Each covariate $X_{ij}(1\leq i \leq N, 1\leq j \leq d)$ is generated from a standard norm distribution $N(0,1)$. The corresponding coefficients for $X_{i}$ are set to be $\beta=(1,2,1)^{\top}$. We then generate the response $Y_i (1\leq i \leq N)$ from a Bernoulli distribution with the probability given as,
\begin{equation}
P(Y_i=1|X_i,\beta)=\frac{\exp(X_{i}^{\top}\beta)}{1+\exp(X_{i}^{\top}\beta)}. \nonumber
\end{equation}

{\sc Example 2. (Poisson Regression)} The poisson regression is used to model count responses. We also consider $d = 3$ exogenous covariates $X_{i}$ in this example, each of which is generated from a uniform distribution $U(0,1)$. The corresponding coefficients $\beta=(1,-1,-0.5)^{\top}$. Then, the response $Y_i (1\leq i \leq N)$ is generated from a Poisson distribution given as
\begin{equation}
P(Y_i|X_i,\beta)=\frac{\lambda_i^{Y_i}}{Y_i!}\exp(\lambda_i), {\text{where }} \lambda_i=\exp(X_{i}^{\top}\beta). \nonumber
\end{equation}

To verify the performance of the one-step estimator, we consider two settings of storing strategies in each example.

{\sc Case 1. (Randomly Distributed)} In this case, the data $\{(X_{i}, Y_i), 1\leq i \leq N\}$ are randomly distributed across the $K$ Workers.

{\sc Case 2. (Nonandomly Distributed)} To make the whole dataset nonrandomly distributed across the $K$ Workers, we conduct the following process. Define $Z_i=\sum_{j=1}^{d}X_{ij}$ as the summation of covariates. Given $\{Z_i, 1\leq i \leq N\}$, define $Z_{(i)}$ as the $i$th order statistics, i.e, $Z_{(1)} \leq Z_{(2)} \leq ... \leq Z_{(N)}$. Then, we store the $(i)$-th observation $(X_{(i)}, Y_{(i)})$ in the $\tau_{i}$-th Worker, where $\tau_{i}=[iK/N]+1$. By this way, the storing place of each observation is related to its covariates, and thus the whole dataset is distributed nonrandomly.

With the pairwise combinations of different examples and cases, we have four simulation scenarios in total. They are, respectively, {\sc{Logistic}} $+$ {\sc Random}, {\sc{Logistic}} $+$ {\sc Nonrandom}, {\sc{Poisson}} $+$ {\sc Random}, and {\sc{Poisson}} $+$ {\sc Nonrandom}. In each scenario, we repeat the experiment for $B=500$ times.

\csubsection{Performance Measurement}

For a reliable evaluation, we take different measures for comparison. For the randomly distributed case, we compare our proposed one-step estimator with: (a) the global (GO) estimator based on the whole sample, (b) the one-shot (OS) estimator, and (c) the communication-efficient surrogate likelihood (CSL) estimator \citep{jordan2018communication}. In the meanwhile, we also report the results for the pilot estimator. For the nonrandomly distributed case, we omit the CSL estimator, since its inference algorithm does not work well with nonrandomly distributed data. In addition, given the pilot sample size $n$ would definitely influence the performance of the one-step estimator, we consider different sizes of the pilot sample. Specifically, define $p=n/N$ as the pilot percentage and we take $p=(0.05,0.1,0.2)$ for illustration.

For one particular method (i.e., GO, OS, CSL, Pilot, and One-Step), we define $\hat{\beta}^{(b)}=(\hat{\beta}_{j}^{(b)})_{j=1}^{d}$ as the estimator in the $b$-th replication. Then, to evaluate the estimation efficiency of each estimator, we define the root mean squared error (RMSE) for $\hat{\beta}_{j}$, i.e., $\text{RMSE}(\hat{\beta}_{j})=\sqrt{B^{-1}\sum_{b=1}^{B}(\hat{\beta}_{j}^{(b)}-\beta_j)^2}$. Finally, we compute the averaged RMSE of all covariates as the final performance measure, i.e., $\text{ARMSE}=d^{-1}\sum_{j=1}^{d}\text{RMSE}(\hat{\beta}_{j})$.

\csubsection{Simulation Results}

Table \ref{t:case1} presents the simulation results for estimation performance under the logistic regression for randomly, and nonrandomly distributed cases, respectively. The corresponding results under the poisson regression have similar patterns, which are shown in Table \ref{t:case2}. From the simulation results in randomly distributed cases, we can draw the following conclusions. First, when data are randomly distributed, both the one-shot estimator and CSL estimator can achieve similar estimation performance with the global one. Also, these three estimators are $\sqrt{N}$-consistent, in the sense that the ARMSE values are close to zero as $N \rightarrow \infty$. Second, the ARMSE values of both pilot estimator and one-step estimator decrease as the pilot percentage or sample size $N$ increase. This finding verifies that the pilot estimator and one-step estimator are $\sqrt{n}$-consistent, in accordance with Theorem 1. Third, although the pilot estimator does not work well when compared with the global one, the one-step estimator performs much better. It can achieve similar ARMSE with the global estimator under a small pilot percentage when $N$ is large. In fact, Theorem 1 shows that when $n^2 \gg N$, the one-step estimator is statistically efficient as the global one. Last, in the randomly distributed case, the estimation performance of different estimators are not related with the number of Workers.

Then, we focus on the simulation results in nonrandomly distributed cases shown in Table \ref{t:case1} and Table \ref{t:case2}. It is notable that, when data are nonrandomly distributed across different Workers, the story of one-shot estimator has totally changed. Specifically, the ARMSE of one-shot estimator is much larger than the global one. In addition, when the number of Workers increases, the data distributed on each Worker become more heterogenous, and the resulted one-shot estimator performs worse. On the contrary, the estimation performance of one-step estimator is similarly with that in the randomly distributed case. This suggests that the statistical efficiency of one-step estimator remains comparable with that of the global one.

\csubsection{Simulation for Likelihood Ratio Tests}

In this subsection, we demonstrate the finite sample performance of likelihood ratio test using one-step estimator. Assume the whole dataset contain $N=(10,20,50)\times 10^{3}$ observations, which are stored in $K=5$ Workers. The examples of logistic regression and poisson regression are both considered, for randomly distributed and nonrandomly distributed cases. For illustration purpose, we only consider two covariates $(X_{i1},X_{i2})(1\leq i \leq N)$, where $X_{i1}=1$ and $X_{i2}$ is independently and identically generated from a uniform distribution $U(0,1)$. The corresponding coefficients for $(X_{i1},X_{i2})$ are $\beta_1$ and $\beta_2$. In the logistic regression, we set $\beta_1=0.2$, and the null hypothesis is $H_0:\beta_2=0$ versus the alternative one $H_1:\beta_2\neq0$. In the poisson regression, we set $\beta_1=0.5$ and also test whether $\beta_2$ equals zero or not.

For each hypothesis testing problem, we conduct a variety of likelihood ratio tests using: (a) the global estimator, (b) the one-shot estimator, (c) the pilot estimator, and (d) the one-step estimator. For $R=$  GO,  Pilot and One-Step estimators, we first obtain $\hat{\beta}_{\text{null}}^{R}=(\hat{\beta}_{\text{null},1}^{R})$ and $\hat{\beta}_{\text{alt}}^{R}=(\hat{\beta}_{\text{alt},1}^{R},\hat{\beta}_{\text{alt},2}^{R})$ under $H_0$ and $H_1$, respectively. Then the likelihood ratio test is applied. Under Theorem 2, the test statistics $\Delta_{R}=2\ln \lambda_{R}(Y)$ follows a $\chi^2(1)$ distribution. For the one-shot estimator, we first use observations stored on the $k$-th Worker to obtain $\hat{\beta}_{\text{null}}^{\text{OS},k}$ and $\hat{\beta}_\text{alt}^{\text{OS},k}$ under $H_0$ and $H_1$, respectively. Then, we summarize the information from all $K$ Workers and calculate the test statistic as
$\lambda_{OS}(Y)=\left\{\prod_{k=1}^K\prod_{i \in \mS_k}p(Y_i,\hat{\beta}_{\text{alt}}^{\text{OS},k})\right\}/\left\{\prod_{k=1}^K\prod_{i \in \mS_k}p(Y_i,\hat{\beta}_{\text{null}}^{\text{OS},k})\right\}$.
It can be easily derived that $\Delta_{\text{OS}}=2\ln \lambda_{OS}(Y)$ follows a $\chi^2(K)$ distribution.

To evaluate the testing performance, we set the significance level $\alpha=0.05$. Define $\Delta_{R}^{(b)}$ as the test statistics ($R=$ GO, OS, Pilot, One-Step) in the $b$-th replication for $1\leq b\leq B$, where $B$ is the number of replications. Then, define an indicator measure $I_{R}^{(b)}=I\{\Delta_{R}^{(b)}>\chi^2_{1-\alpha}(1)\}$ for using global, pilot and one-step estimators, and $I^{(b)}_{\text{OS}}=I\{\Delta_{\text{OS}}^{(b)}>\chi^2_{1-\alpha}(K)\}$ for using one-shot estimator. For a total of $B=500$ replications, we can calculate the empirical rejection probability as $\text{ERP}=B^{-1}\sum_{b=1}^{B}I_{R}^{(b)}$. Therefore, the $\text{ERP}$ is the empirical size under the null hypothesis $H_0$, while it is the empirical power under the alternative hypothesis $H_1$. The empirical size and power provide overall measures for assessing the performance of the likelihood ratio test.

Simulation results for likelihood ratio tests under the logistic regression and poisson regression are shown in Table \ref{t:LRT1} and Table \ref{t:LRT2}, respectively. We find that, all likelihood ratio tests control the size very well, which are around the nominal level 5\%. However, they perform differently in empirical power. Specifically, the likelihood ratio tests using global estimator and one-step estimator result in similar empirical power, which are the best of all. The bad performance for likelihood ratio test using pilot estimator is mainly due to its smaller sample size $n$. As for the OS method, it has obtained lower empirical power than tests using global estimator and one-step estimator. The performance of OS method is even worse than the pilot estimator when data are nonrandomly distributed among Workers. Finally, we find the pilot percentage does not affect the empirical size and power much for test using one-step estimator. This finding implies that a small pilot percentage is enough in practice for conducting likelihood ratio test using one-step estimator.

\csection{Real Data Analysis}

For illustration purpose, we elaborate the application of our proposed method on an American Airline dataset. This is a public dataset, which can be downloaded from \url{http://stat-computing.org/dataexpo/2009/}. It contains flight arrival and departure information for all commercial flights from 1987 to 2007 in America. There are a total of nearly 120 million records and take up 12GB on a hard drive. Each record contains information such as the delayed status, the actual and scheduled departure or arrival time, the flying distance, and so on. We place this dataset on a Spark-on-YARN cluster, which is set up on Aliyun Cloud Server, a industrial-level service for distributed systems (\url{https://www.aliyun.com/product/emapreduce}). The Spark system consists of one Master and seven Workers. For convenience, all records in every three years are stored in one Worker.

\begin{figure}[h]
	\centerline{\includegraphics[width=3.5in]{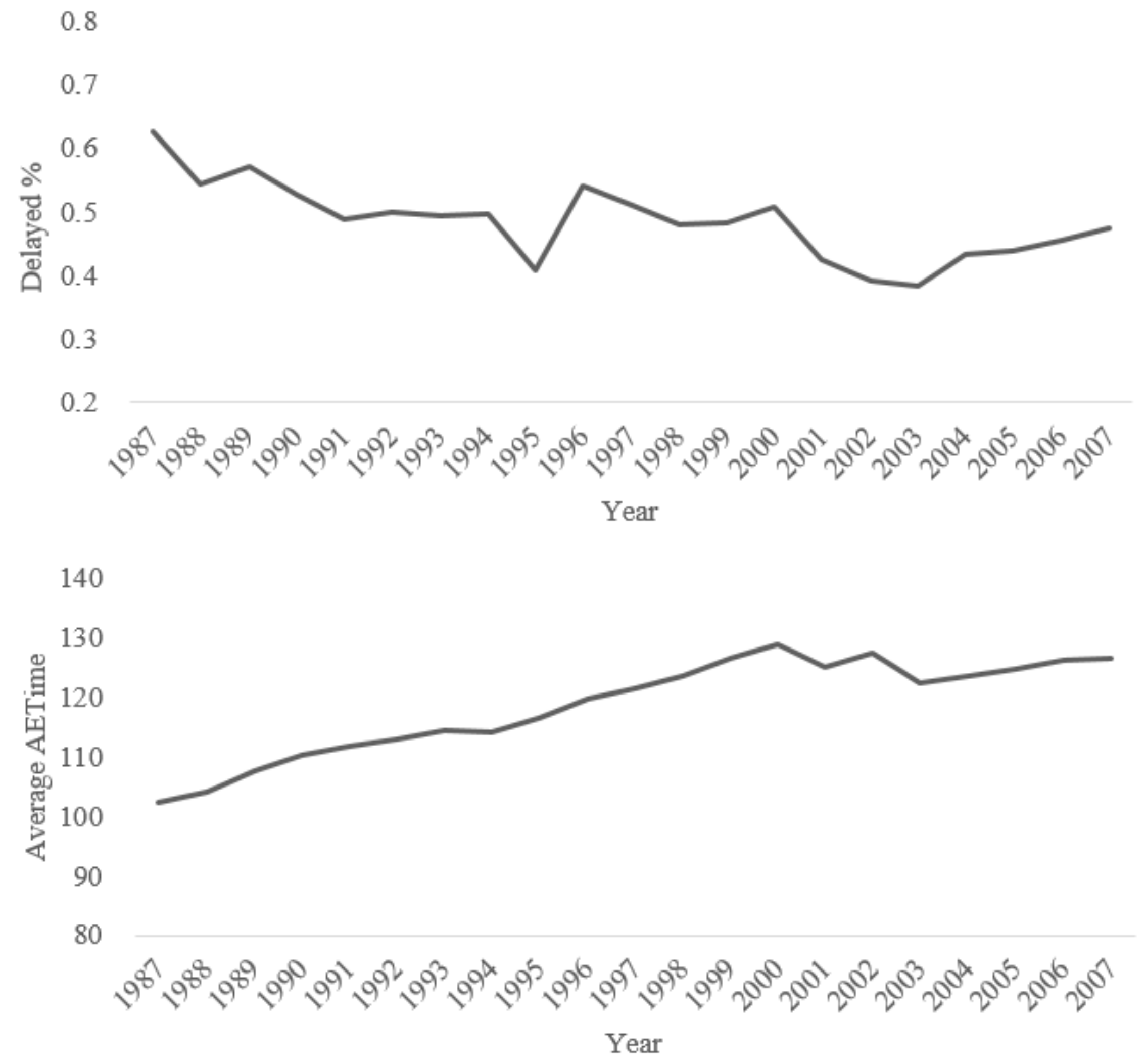}}
	\caption{The time trend of delayed percentages of flights (the top) and the average of actual elapsed time (the bottom) from 1987 to 2007. }
	\label{f:plot}
\end{figure}

The research goal of this dataset is to investigate factors that can influence the delayed status of flight. To this end, we take ``Delayed", indicating whether the flight is delayed or not, as the response, and consider four continuous variables as covariates for illustration purpose. The detailed information of each variable is shown in Table \ref{t:description}. To investigate the nonrandomly distributed nature of this dataset, Figure \ref{f:plot} presents the delayed percentages of flights and the average of actual elapsed time in each year from 1987 to 2007. It is obvious that, the two variables behave differently in different years. The delayed percentages of flights in most years vary between 0.4 to 0.6. We can also observe a decreasing trend in the delayed percentages. On the contrary, the averages of actual elapsed time are rising continuously. These findings imply the data stored by every three years have the nonrandomly distributed issue.

To address this issue, we establish a logistic regression and use the proposed one-step method for estimation. The pilot percentage is set as 0.001.  For comparison purpose, we also obtain the global estimator and the one-shot estimator based on the whole dataset. The likelihood ratio tests are then applied to investigate the significance for each estimator. Finally, we compute the log-likelihood of each method to evaluate the model performance.

Table \ref{t:results} presents the detailed regression results under the global method, one-shot method and one-step method. It is notable that, in different methods, the p.values of all covariates are smaller than 0.001, suggesting their significant influences to the response. In general, the one-step estimates are more similar with the global ones, especially for the covariates ``ActualElapsedTime" and ``Distance".
As for the model fitting performance evaluated by log-likelihood, the global method has achieved the highest log-likelihood, indicating its best performance of all. It is followed by the one-step method, which is lower than the global method, but much higher than the one-shot method. All these findings verify the better performance of one-step method in handling nonrandomly distributed data.

\csection{CONCLUDING REMARKS}

In this work, we develop a Pseudo-Newton-Raphson algorithm for an efficient estimation of GLM on a distributed system with nonrandomly distributed data. By only using two rounds of Master-and-Worker communication, this algorithm is computationally efficient in both communication cost and storage usage. The final one-step estimator is also proved as statistically efficient as the global estimator under some technical conditions. Based on the one-step estimator, a likelihood ratio test for hypothesis testing is also developed. The performances of the estimator and the corresponding likelihood ratio test are elaborated by both simulation studies and a real American airline dataset.

To conclude this work, we consider some directions for future study. First, the covariates in large-scale datasets are typically of high dimensionality. Then how to conduct feature selection based on the Pseudo-Newton-Raphson algorithm is worth of consideration. Second, in the first step, we have to collect a sample of size $n$ from Workers to the Master. To further reduce the communication cost in transferring data, some sufficient statistics are worth of investigation. Last, the proposed one-step method can be further extended to solve model estimation problems using Newton-Raphson algorithms.

\newpage
\scsection{Appendices }
\setcounter{lemma}{0}
\setcounter{theorem}{0}
\renewcommand{\theequation}{A.\arabic{equation}}
\setcounter{equation}{0}
\setcounter{table}{0}

\scsubsection{Appendix A. Basic Matrices and The Estimation Procedure}

Define $\dot\ell(\beta)$ and $\ddot\ell(\beta)$ to be the first and second order derivatives of $\ell_f(\beta)$ respectively.
By Theorem 2 in \cite{Wedderburn:1974}, we have $\partial \ell(Y_i,\mu_i)/\partial \mu_i=(Y-\mu_i)/V(\mu_i)$. Furthermore, it could be proved that,
\beqrs
&&\bE\left(\frac{\partial \ell_i}{\partial \mu_i}\right)=\bE\left(\frac{\partial \ell_i}{\partial \beta_j}\right)=0,\\
&&\bE\left(\frac{\partial \ell_i}{\partial \mu}\right)^2=-\bE\left(\frac{\partial ^2\ell_i}{\partial \mu_i^2}\right)=\frac{1}{V(\mu_i)},\\
&&\bE\left(\frac{\partial \ell_i}{\partial \beta}\frac{\partial \ell_i}{\partial \beta^\top}\right)=\bE\left(\frac{\partial^2 \ell_i}{\partial \beta\partial \beta^\top}\right)=\frac{1}{V(\mu_i)}\frac{\partial \mu_i}{\partial \beta}\frac{\partial \mu_i}{\partial \beta^\top}
\eeqrs
where the $\bE$ denotes the expectation given the covariates $X$. It could be calculated that,
 \beqrs
 \ddot\ell(\beta)=\frac{1}{N}\sum_{i=1}^N\Big[(Y_i-\mu_i)\frac{\partial}{\partial \beta^\top}\Big\{\frac{1}{V(\mu_i)}\frac{\partial \mu_i}{\partial\beta}\Big\}-\frac{1}{V(\mu_i)}\frac{\partial \mu_i}{\partial\beta}\frac{\partial \mu_i}{\partial\beta^\top}\Big].
 \eeqrs
 Thus it could be verified that,
\beqr
\frac{1}{N}\dot\ell(\beta)&=&\frac{1}{N}\sum_{i=1}^N\frac{(Y_i-\mu_i)}{V(\mu_{i})}\frac{\partial \mu_i}{\partial \beta},\n\\
\frac{1}{N}\bE\ddot\ell(\beta)&=&\frac{1}{N}\sum_{i=1}^N\frac{1}{V(\mu_{i})}\frac{\partial \mu_i}{\partial \beta}\frac{\partial \mu_i}{\partial \beta^\top}.\label{eq:condsec}
\eeqr
By Taylor expansion, $\hat\beta_M=\beta_0+\{N^{-1}\ddot\ell(\beta_0)\}^{-1}\{N^{-1}\dot\ell(\beta_0)\}+o_p(N^{-1/2})$. We replace  $N^{-1}\ddot\ell(\beta_0)$ with $N^{-1}\bE\ddot\ell(\beta_0)$ in (\ref{eq:condsec}) and obtain (\ref{eq:onestep}).

\scsubsection{Appendix B. Proof of Theorem 1}

To prove the theorem, it is sufficient to show that $\hat\beta_M-\hat\beta_o=o_p(N^{-1/2})$. Based on (\ref{eq:onestep}) and (\ref{eq:onestep1}), it could be verified that, $\hat\beta_M-\hat\beta_o=S_1+S_2+S_3+o_p(N^{-1/2})$, where
\beqrs
S_1&=&\beta_0-\hat\beta_p+\Big\{\frac{1}{N}\sum_{i=1}^N\frac{1}{V(\mu_{i0})}\dot\mu_{i}\dot\mu_{i}^\top\Big\}^{-1}\Big[\frac{1}{N}\sum_{i=1}^N\Big\{\frac{(Y_i-\mu_{i0})}{V(\mu_{i0})}\dot\mu_{i}-\frac{(Y_i-\hat\mu_{pi})}{V(\hat\mu_{pi})}\dot\mu_{pi}\Big\}\Big],\\
S_2&=&\Big[\Big\{\frac{1}{N}\sum_{i=1}^N\frac{1}{V(\mu_{i0})}\dot\mu_{i}\dot\mu_{i}^\top\Big\}^{-1}-\Big\{\frac{1}{N}\sum_{i=1}^N\frac{1}{V(\hat\mu_{pi})}\dot\mu_{pi}\dot\mu_{pi}^\top\Big\}^{-1}\Big]\frac{1}{N}\sum_{i=1}^N\frac{(Y_i-\mu_{i0})}{V(\mu_{i0})}\dot\mu_{i}\\
S_3&=&\Big[\Big\{\frac{1}{N}\sum_{i=1}^N\frac{1}{V(\mu_{i0})}\dot\mu_{i}\dot\mu_{i}^\top\Big\}^{-1}-\Big\{\frac{1}{N}\sum_{i=1}^N\frac{1}{V(\hat\mu_{pi})}\dot\mu_{pi}\dot\mu_{pi}^\top\Big\}^{-1}\Big]\Big[\frac{1}{N}\sum_{i=1}^N\Big\{\frac{(Y_i-\hat\mu_{pi})}{V(\hat\mu_{pi})}\dot\mu_{pi}\\
&&~~~~~~~~~~~~~~~~-\frac{(Y_i-\mu_{i0})}{V(\mu_{i0})}\dot\mu_{i}\Big\}\Big].
\eeqrs

First, we consider $S_1$.
 Define $\dot\ell_j(\beta)=N^{-1}\sum_iV(\mu_i)^{-1}(Y_i-\mu_i)\dot \mu_{ij}$ is the $j$th element of $\dot\ell(\beta)$, which is the first order derivative of $\ell_f(\beta)$ with respect to $\beta$. It is remarkable that $\hat\beta_p-\beta_0=O_p(n^{-1/2})$. Then we conduct Taylor expansion on $\dot\ell_j(\beta)$ with respect to $\beta$ at point $\beta_0$ and one could obtained,
$\dot\ell_j(\hat\beta_p)=\dot\ell_j(\beta_0)+\partial \dot\ell_j(\hat\beta_p)/\partial \beta(\hat\beta_p-\beta_0)+O_p(n^{-1})$,
where, $\partial \dot\ell_j(\hat\beta_p)/\partial \beta$ is the $j$th element of $\partial \dot\ell(\beta)/\partial \beta|_{\beta=\hat\beta_p}$.
 It could be verified that $\dot\ell(\beta_0)-\dot\ell(\hat\beta_p)=-\ddot\ell(\beta_0)(\hat\beta_p-\beta_0)+O_p(n^{-1})$, where
 \beqrs
 \ddot\ell(\beta_0)=\frac{1}{N}\sum_{i=1}^N\Big[(Y_i-\mu_i)\frac{\partial}{\partial \beta^\top}\Big\{\frac{1}{V(\mu_i)}\frac{\partial \mu_i}{\partial\beta}\Big\}\Big|_{\beta=\beta_0}-\frac{1}{V(\mu_i)}\frac{\partial \mu_i}{\partial\beta}\frac{\partial \mu_i}{\partial\beta^\top}\Big|_{\beta=\beta_0}\Big].
 \eeqrs
 By central limit theorem, $N^{-1}\sum_i\big[(Y_i-\mu_i)\partial\{V(\mu_i)^{-1}\partial \mu_i/\partial \beta\}/\partial \beta^\top\big]|_{\beta=\beta_0}$ is $O_p(N^{-1/2})$. Thus $\dot\ell(\beta_0)-\dot\ell(\hat\beta_p)=N^{-1}\sum_{i}V(\mu_{i0})^{-1}\dot\mu_i\dot\mu_i^\top(\hat\beta_p-\beta_0)+O_p(n^{-1})$. Since $n^2/N\rightarrow \infty$, thus it could be verified that $S_1=o_p(N^{-1/2})$.
 %
%

 Second, we prove that $\Big\{N^{-1}\sum_{i=1}^N V(\mu_{i0})^{-1}\dot\mu_{i}\dot\mu_{i}^\top\Big\}^{-1}-\Big\{N^{-1}\sum_{i=1}^N V(\hat\mu_{pi})^{-1}\dot\mu_{pi}\dot\mu_{pi}^\top\Big\}^{-1}$ is $o_p(1)$. Furthermore, we define $M_N(\beta_0)=N^{-1}\sum_{i=1}^N V(\mu_{i0})^{-1}\dot\mu_{i}\dot\mu_{i}^\top$ and $M_N(\hat\beta_p)=N^{-1}\sum_{i=1}^N V(\hat\mu_{pi})^{-1}\dot\mu_{pi}\dot\mu_{pi}^\top$. Since the dimension of the matrix is fixed to be $d$, it is sufficient to show that each of its element to be $o_p(1)$. Consider $M_N(\beta_0)^{-1}-M_N(\hat\beta_p)^{-1}=M_N(\beta_0)^{-1}\big\{M_N(\hat\beta_p)-M_N(\beta_0)\big\}M_N(\hat\beta_p)^{-1}$. It could be verified that $M_N(\beta_0)^{-1}\rightarrow_p M^{-1}$ and $M_N(\hat\beta_p)^{-1}\rightarrow_p M^{-1}$. Thus it is sufficient to show that each element of $M_N(\hat\beta_p)-M_N(\beta_0)$ is $o_p(1)$. Define $f_{N,j_1j_2}(\beta)=N^{-1}\sum_{i=1}^N V(\mu_{i})^{-1}(\partial \mu_i/\partial\beta_{j_1})(\partial \mu_i/\partial\beta_{j_2})$.
 Thus we have, $f_{N,j_1j_2}(\hat\beta_p)-f_{N,j_1j_2}(\beta_0)=\{\partial f_{N,j_1j_2}(\beta)/\partial \beta|_{\beta=\beta_0}\}(\hat\beta_p-\beta_0)+O_p(n^{-1})$. As a result, by Law of Large Numbers, it could be concluded that $\partial f_{N,j_1j_2}(\beta)/\partial \beta|_{\beta=\beta_0}\rightarrow_p E[\partial\{ V(\mu_{i})^{-1}(\partial \mu_i/\partial\beta_{j_1})(\partial \mu_i/\partial\beta_{j_2})\}/\partial \beta|_{\beta=\beta_0}]$. Thus, we have $f_{N,j_1j_2}(\hat\beta_p)-f_{N,j_1j_2}(\beta_0)=O_p(n^{-1/2})=o_p(1)$.  Then, by Central Limit Theorem,  $N^{-1}\sum_{i=1}^N(Y_i-\mu_{i0})V(\mu_{i0})^{-1}\dot\mu_{i}=O_p(N^{-1/2})$. Thus $S_2=o_p(1)O_p(N^{-1/2})=o_p(N^{-1/2})$. And $S_3=O_p(n^{-1/2})O_p(n^{-1/2})=O(n^{-1})=o_p(N^{-1/2})$. This completes the proof of the theorem.


 \scsubsection{Appendix C. Proof of Theorem 2}

In this section, we prove Theorem 2 in a more general case. Assume that we would like to test the null hypothesis H$_{0}:\beta_s=\beta_{s0}$ versus the alternative one H$_1:\beta_s\neq\beta_{s0}$. Thus we could obtain the MLEs under both the null hypothesis and the alternative one, which are defined as $\hat\beta_M$ and $\hat\beta_{0,M}$ respectively. Let $\hat\beta_{0,M}=(\beta_{s0}^\top,\hat\beta_{cM}^\top)^\top$, where $\hat\beta_{cM}$ is the MLE of $\beta_c$ under the alternative hypothesis. Furthermore, we define $\lambda_M(Y)=\{\prod_{i=1}^N p(Y_i,\hat\beta_{0,M})\}^{-1}\{\prod_{i=1}^N p(Y_i,\hat\beta_M)\}.$  Similarly we could define $\lambda_o(Y)=\{\prod_{i=1}^N p(Y_i,\hat\beta_{0,o})\}^{-1}\{\prod_{i=1}^N p(Y_i,\hat\beta_o)\}.$ We are going to prove that $2\ln \lambda_o(Y)$  asymptotically follows a $\chi^2(s)$ distribution, where $s$ is the degree of freedom. It is remarkable that Theorem 2 is a special case of this conclusion.

We only need to prove that $2\ln \lambda_M(Y)-2\ln \lambda_o(Y)=o_p(1)$. It could be expressed that,
 $2\ln \lambda_M(Y)=2\sum_{i=1}^N\ell(Y_i,\hat\beta_M)-2\sum_{i=1}^N\ell(Y_i,\hat\beta_{0,M})$, and
  $2\ln \lambda_o(Y)=2\sum_{i=1}^N\ell(Y_i,\hat\beta_o)-2\sum_{i=1}^N\ell(Y_i,\hat\beta_{0,o})$.
Thus we only need to prove that,
 \beqr
 \ell (\hat\beta_M)-\ell(\hat\beta_o)&=&o_p(1)\label{eq:op1}\\
 \ell (\hat\beta_{0,M})-\ell(\hat\beta_{0,o})&=&o_p(1).\label{eq:op2}
 \eeqr
  By the proof of Theorem \ref{thm1}, we have $\hat\beta_M-\hat\beta_o=o_p(N^{-1/2})$. Similarly, it could be verified that $\hat\beta_{0,M}-\hat\beta_{0,o}=o_p(N^{-1/2})$. Since the proof of (\ref{eq:op1}) and (\ref{eq:op1}), we only prove (\ref{eq:op1}). By Taylor expansion, we have
  \beqrs
  \ell(\hat\beta_o)=\ell(\hat\beta_M)+(\hat\beta_o-\hat\beta_M)^\top\dot \ell(\hat\beta_M)+(\hat\beta_o-\hat\beta_M)^\top\ddot\ell(\beta^*)(\hat\beta_o-\hat\beta_M),
  \eeqrs
  where $\beta^*$ is between $\hat\beta_o$ and $\hat\beta_M$.
  Since $\dot \ell(\hat\beta_M)=0$, we have $\ell(\hat\beta_o)-\ell(\hat\beta_M)=\sqrt{N}(\hat\beta_o-\hat\beta_M)^\top \{N^{-1}\ddot\ell(\beta^*)\}\sqrt{N}(\hat\beta_o-\hat\beta_M)$. Define $C(\beta^*)=(c_{k_1k_2}(\beta^*))=-N^{-1}\ddot\ell(\beta^*)$. Thus,
  \beqrs
  c_{k_1k_2}(\beta^*)&=&c_{k_1k_2}(\hat\beta_M)+(\beta^*-\hat\beta_M)^\top\frac{\partial c_{ij}(\beta)}{\partial \beta}\Big|_{\beta=\hat\beta_M-\gamma(\hat\beta_M-\beta^*)}\\
  &\leq&c_{k_1k_2}(\hat\beta_M)+\big(\sum_{l=1}^d|\beta_l^*-\hat\beta_{M,l}|\big)N^{-1}\sum_{i=1}^NM(Y_i),
  \eeqrs
  where $0\leq \gamma\leq 1$. Since $\beta_l^*-\hat\beta_{M,l}=o_p(1)$, and $N^{-1}\sum_{i=1}^NM(Y_i)\rightarrow_p EM(Y_i)<K$. Thus $c_{k_1k_2}(\beta^*)-c_{k_1k_2}(\hat\beta_M)=o_p(1)$. Since $c_{k_1k_2}(\hat\beta_M)\rightarrow \pM_{k_1k_2}(\beta_0)$, we have $c_{k_1k_2}(\beta^*)\rightarrow \pM_{k_1k_2}(\beta_0)$. As a result, $C(\beta^*)\rightarrow \pM(\beta_0)$. Together with $\sqrt{N}(\hat\beta_o-\hat\beta_M)=o_p(1)$, we have $\ell (\hat\beta_M)-\ell(\hat\beta_o)=o_p(1)$. This completes the proof.

\newpage

\bibliographystyle{asa}
\bibliography{danyang}

\newpage
\begin{table}[]
\caption{Simulation Results for Estimation Performance under Logistic Regression in Randomly and Nonrandomly Distributed Cases.}\label{t:case1}
\begin{center}
\renewcommand\arraystretch{1}
\begin{threeparttable}[b]
\begin{tabular}{cc|ccccccccccc}
\hline
\hline
\multirow{2}{*}{$K$} &\multirow{2}{*}{$p\%$} & \multicolumn{5}{c}{Random} & & \multicolumn{4}{c}{Nonrandom}\\
\cline{3-7}
\cline{9-12}
& &GO	&OS &CSL 		&Pilot	&One-Step & &GO	&OS		&Pilot	&One-Step\\
\hline
& &\multicolumn{10}{c}{\multirow{2}*{$N=10,000$}}\\
& &\multicolumn{10}{c}{}\\
\multirow{3}{*}{2}	&5	&\multirow{3}{*}{0.034}	&\multirow{3}{*}{0.034}	&\multirow{3}{*}{0.034}	&0.147	&0.049	&	&\multirow{3}{*}{0.034}	&\multirow{3}{*}{0.049}	&0.156	&0.050	\\
	&10	&	&	&	&0.112	&0.039	&	&	&	&0.108	&0.037	\\
	&20	&	&	&	&0.078	&0.033	&	&	&	&0.075	&0.034	\\
& &\multicolumn{10}{c}{}\\
\multirow{3}{*}{5}	&5	&\multirow{3}{*}{0.035}	&\multirow{3}{*}{0.036}	&\multirow{3}{*}{0.035}	&0.167	&0.061	&	&\multirow{3}{*}{0.035}	&\multirow{3}{*}{0.184}	&0.159	&0.053	\\
	&10	&	&	&	&0.112	&0.039	&	&	&	&0.107	&0.038	\\
	&20	&	&	&	&0.078	&0.035	&	&	&	&0.077	&0.034	\\
& &\multicolumn{10}{c}{}\\
\multirow{3}{*}{10}	&5	&\multirow{3}{*}{0.035}	&\multirow{3}{*}{0.038}	&\multirow{3}{*}{0.035}	&0.161	&0.053	&	&\multirow{3}{*}{0.034}	&\multirow{3}{*}{6.096}	&0.16	&0.065	\\
	&10	&	&	&	&0.107	&0.039	&	&	&	&0.105	&0.038	\\
	&20	&	&	&	&0.076	&0.034	&	&	&	&0.081	&0.034	\\
\hline
& &\multicolumn{10}{c}{\multirow{2}*{$N=20,000$}}\\
& &\multicolumn{10}{c}{}\\
\multirow{3}{*}{2}	&5	&\multirow{3}{*}{0.024}	&\multirow{3}{*}{0.024}	&\multirow{3}{*}{0.024}	&0.111	&0.031	&	&\multirow{3}{*}{0.025}	&\multirow{3}{*}{0.034}	&0.112	&0.030	\\
	&10	&	&	&	&0.079	&0.024	&	&	&	&0.078	&0.025	\\
	&20	&	&	&	&0.055	&0.024	&	&	&	&0.055	&0.025	\\
& &\multicolumn{10}{c}{}\\
\multirow{3}{*}{5}	&5	&\multirow{3}{*}{0.024}	&\multirow{3}{*}{0.025}	&\multirow{3}{*}{0.024}	&0.107	&0.031	&	&\multirow{3}{*}{0.025}	&\multirow{3}{*}{0.118}	&0.11	&0.029	\\
	&10	&	&	&	&0.076	&0.026	&	&	&	&0.075	&0.025	\\
	&20	&	&	&	&0.055	&0.024	&	&	&	&0.054	&0.024	\\
& &\multicolumn{10}{c}{}\\
\multirow{3}{*}{10}	&5	&\multirow{3}{*}{0.025}	&\multirow{3}{*}{0.026}	&\multirow{3}{*}{0.025}	&0.107	&0.030	&	&\multirow{3}{*}{0.024}	&\multirow{3}{*}{0.777}	&0.112	&0.032	\\
	&10	&	&	&	&0.080	&0.025	&	&	&	&0.078	&0.026	\\
	&20	&	&	&	&0.053	&0.025	&	&	&	&0.049	&0.024	\\
\hline
& &\multicolumn{10}{c}{\multirow{2}*{$N=100,000$}}\\
& &\multicolumn{10}{c}{}\\
\multirow{3}{*}{2}	&5	&\multirow{3}{*}{0.011}	&\multirow{3}{*}{0.011}	&\multirow{3}{*}{0.011}	&0.050	&0.011	&	&\multirow{3}{*}{0.010}	&\multirow{3}{*}{0.015}	&0.048	&0.011	\\
	&10	&	&	&	&0.034	&0.011	&	&	&	&0.033	&0.011	\\
	&20	&	&	&	&0.023	&0.011	&	&	&	&0.024	&0.010	\\
& &\multicolumn{10}{c}{}\\
\multirow{3}{*}{5}	&5	&\multirow{3}{*}{0.011}	&\multirow{3}{*}{0.011}	&\multirow{3}{*}{0.011}	&0.048	&0.011	&	&\multirow{3}{*}{0.011}	&\multirow{3}{*}{0.052}	&0.049	&0.012	\\
	&10	&	&	&	&0.034	&0.011	&	&	&	&0.033	&0.011	\\
	&20	&	&	&	&0.022	&0.011	&	&	&	&0.024	&0.011	\\
& &\multicolumn{10}{c}{}\\
\multirow{3}{*}{10}	&5	&\multirow{3}{*}{0.011}	&\multirow{3}{*}{0.011}	&\multirow{3}{*}{0.011}	&0.049	&0.011	&	&\multirow{3}{*}{0.011}	&\multirow{3}{*}{0.133}	&0.05	&0.011	\\
	&10	&	&	&	&0.032	&0.011	&	&	&	&0.035	&0.011	\\
	&20	&	&	&	&0.025	&0.011	&	&	&	&0.024	&0.010	\\
\hline
\hline
\end{tabular}
\end{threeparttable}
\end{center}
\end{table}

\begin{table}[]
\caption{Simulation Results for Estimation Performance under Poisson Regression in Randomly and Nonrandomly Distributed Cases.}\label{t:case2}
\begin{center}
\renewcommand\arraystretch{1}
\begin{threeparttable}[b]
\begin{tabular}{cc|ccccccccccc}
\hline
\hline
\multirow{2}{*}{$K$} &\multirow{2}{*}{$p\%$} & \multicolumn{5}{c}{Random} & & \multicolumn{4}{c}{Nonrandom}\\
\cline{3-7}
\cline{9-12}
& &GO	&OS &CSL 		&Pilot	&One-Step & &GO	&OS		&Pilot	&One-Step\\
\hline
& &\multicolumn{10}{c}{\multirow{2}*{$N=10,000$}}\\
& &\multicolumn{10}{c}{}\\
\multirow{3}{*}{2}	&5	&\multirow{3}{*}{0.037}	&\multirow{3}{*}{0.037}	&\multirow{3}{*}{0.037}	&0.175	&0.038	&	&\multirow{3}{*}{0.035}	&\multirow{3}{*}{0.049}	&0.166	&0.037	\\
	&10	&	&	&	&0.121	&0.037	&	&	&	&0.115	&0.036	\\
	&20	&	&	&	&0.084	&0.037	&	&	&	&0.085	&0.036	\\
& &\multicolumn{10}{c}{}\\
\multirow{3}{*}{5}	&5	&\multirow{3}{*}{0.037}	&\multirow{3}{*}{0.037}	&\multirow{3}{*}{0.037}	&0.165	&0.038	&	&\multirow{3}{*}{0.035}	&\multirow{3}{*}{0.127}	&0.169	&0.037	\\
	&10	&	&	&	&0.112	&0.038	&	&	&	&0.115	&0.036	\\
	&20	&	&	&	&0.083	&0.037	&	&	&	&0.082	&0.035	\\
& &\multicolumn{10}{c}{}\\
\multirow{3}{*}{10}	&5	&\multirow{3}{*}{0.037}	&\multirow{3}{*}{0.037}	&\multirow{3}{*}{0.037}	&0.167	&0.038	&	&\multirow{3}{*}{0.035}	&\multirow{3}{*}{0.266}	&0.159	&0.037	\\
	&10	&	&	&	&0.116	&0.037	&	&	&	&0.117	&0.036	\\
	&20	&	&	&	&0.084	&0.037	&	&	&	&0.083	&0.035	\\
\hline
& &\multicolumn{10}{c}{\multirow{2}*{$N=20,000$}}\\
& &\multicolumn{10}{c}{}\\
\multirow{3}{*}{2}	&5	&\multirow{3}{*}{0.026}	&\multirow{3}{*}{0.026}	&\multirow{3}{*}{0.026}	&0.128	&0.027	&	&\multirow{3}{*}{0.026}	&\multirow{3}{*}{0.038}	&0.118	&0.027	\\
	&10	&	&	&	&0.083	&0.026	&	&	&	&0.080	&0.026	\\
	&20	&	&	&	&0.063	&0.026	&	&	&	&0.058	&0.026	\\
& &\multicolumn{10}{c}{}\\
\multirow{3}{*}{5}	&5	&\multirow{3}{*}{0.026}	&\multirow{3}{*}{0.026}	&\multirow{3}{*}{0.026}	&0.117	&0.027	&	&\multirow{3}{*}{0.026}	&\multirow{3}{*}{0.084}	&0.117	&0.027	\\
	&10	&	&	&	&0.083	&0.026	&	&	&	&0.084	&0.026	\\
	&20	&	&	&	&0.058	&0.026	&	&	&	&0.057	&0.026	\\
& &\multicolumn{10}{c}{}\\
\multirow{3}{*}{10}	&5	&\multirow{3}{*}{0.026}	&\multirow{3}{*}{0.026}	&\multirow{3}{*}{0.026}	&0.118	&0.027	&	&\multirow{3}{*}{0.026}	&\multirow{3}{*}{0.19}	&0.119	&0.027	\\
	&10	&	&	&	&0.081	&0.026	&	&	&	&0.088	&0.026	\\
	&20	&	&	&	&0.059	&0.026	&	&	&	&0.060	&0.026	\\
\hline
& &\multicolumn{10}{c}{\multirow{2}*{$N=100,000$}}\\
& &\multicolumn{10}{c}{}\\
\multirow{3}{*}{2}	&5	&\multirow{3}{*}{0.012}	&\multirow{3}{*}{0.012}	&\multirow{3}{*}{0.012}	&0.053	&0.012	&	&\multirow{3}{*}{0.011}	&\multirow{3}{*}{0.016}	&0.052	&0.011	\\
	&10	&	&	&	&0.037	&0.012	&	&	&	&0.038	&0.011	\\
	&20	&	&	&	&0.027	&0.012	&	&	&	&0.027	&0.011	\\
& &\multicolumn{10}{c}{}\\
\multirow{3}{*}{5}	&5	&\multirow{3}{*}{0.012}	&\multirow{3}{*}{0.012}	&\multirow{3}{*}{0.012}	&0.050	&0.012	&	&\multirow{3}{*}{0.011}	&\multirow{3}{*}{0.039}	&0.051	&0.011	\\
	&10	&	&	&	&0.037	&0.012	&	&	&	&0.034	&0.011	\\
	&20	&	&	&	&0.026	&0.012	&	&	&	&0.027	&0.011	\\
& &\multicolumn{10}{c}{}\\
\multirow{3}{*}{10}	&5	&\multirow{3}{*}{0.012}	&\multirow{3}{*}{0.012}	&\multirow{3}{*}{0.012}	&0.052	&0.012	&	&\multirow{3}{*}{0.011}	&\multirow{3}{*}{0.073}	&0.053	&0.011	\\
	&10	&	&	&	&0.036	&0.012	&	&	&	&0.037	&0.011	\\
	&20	&	&	&	&0.025	&0.012	&	&	&	&0.025	&0.011	\\
\hline
\hline
\end{tabular}
\end{threeparttable}
\end{center}
\end{table}

\begin{landscape}
\begin{table}[]
\caption{Simulation Results for Likelihood Ratio Test under Logistic Regression.}\label{t:LRT1}
\begin{center}
\renewcommand\arraystretch{1}
\begin{threeparttable}[b]
\begin{tabular}{c|cccccccccccc}
\hline
\hline
\multirow{2}{*}{Storing Strategy} &\multirow{2}{*}{$N$} &\multicolumn{5}{c}{Size} & &\multicolumn{5}{c}{Power} \\
\cline{3-13}
& &GO	&OS	&pilot\%	&Pilot	&One-step & &GO	&OS	&pilot\%	&Pilot	&One-step \\
\hline
\multirow{9}{*}{Random}
&\multirow{3}{*}{10,000}	&\multirow{3}{*}{0.052}	&\multirow{3}{*}{0.054}	&5	&0.050	&0.052	& &\multirow{3}{*}{0.408}	&\multirow{3}{*}{0.226}	&5	&0.050	&0.408	\\
&	&	&	&10	&0.046	&0.052	& &	&	&10	&0.084	&0.408	\\
&	&	&	&20	&0.052	&0.052	& &	&	&20	&0.104	&0.408	\\
\cline{2-13}
&\multirow{3}{*}{20,000}	&\multirow{3}{*}{0.048}	&\multirow{3}{*}{0.046}	&5	&0.040	&0.048	& &\multirow{3}{*}{0.704}	&\multirow{3}{*}{0.422}	&5	&0.094	&0.702	\\
&	&	&	&10	&0.060	&0.048	& &	&	&10	&0.116	&0.702	\\
&	&	&	&20	&0.044	&0.048	& &	&	&20	&0.208	&0.704	\\
\cline{2-13}
&\multirow{3}{*}{50,000}	&\multirow{3}{*}{0.051}	&\multirow{3}{*}{0.05}	&5	&0.046	&0.051	& &\multirow{3}{*}{0.950}	&\multirow{3}{*}{0.846}	&5	&0.134	&0.950	\\
&	&	&	&10	&0.054	&0.051	& &	&	&10	&0.236	&0.950	\\
&	&	&	&20	&0.054	&0.051	& &	&	&20	&0.410	&0.950	\\
\hline
\multirow{9}{*}{Nonrandom}
&\multirow{3}{*}{10,000}	&\multirow{3}{*}{0.051}	&\multirow{3}{*}{0.054}	&5	&0.040	&0.051	& &\multirow{3}{*}{0.388}	&\multirow{3}{*}{0.056}	&5	&0.080	&0.386	\\
&	&	&	&10	&0.056	&0.051	& &	&	&10	&0.084	&0.388	\\
&	&	&	&20	&0.054	&0.051	& &	&	&20	&0.134	&0.388	\\
\cline{2-13}
&\multirow{3}{*}{20,000}	&\multirow{3}{*}{0.048}	&\multirow{3}{*}{0.052}	&5	&0.052	&0.048	& &\multirow{3}{*}{0.690}	&\multirow{3}{*}{0.062}	&5	&0.060	&0.690	\\
&	&	&	&10	&0.052	&0.048	& &	&	&10	&0.094	&0.690	\\
&	&	&	&20	&0.048	&0.048	& &	&	&20	&0.184	&0.690	\\
\cline{2-13}
&\multirow{3}{*}{50,000}	&\multirow{3}{*}{0.052}	&\multirow{3}{*}{0.048}	&5	&0.056	&0.052	& &\multirow{3}{*}{0.966}	&\multirow{3}{*}{0.072}	&5	&0.118	&0.966	\\
&	&	&	&10	&0.056	&0.052	& &	&	&10	&0.252	&0.966	\\
&	&	&	&20	&0.054	&0.052	& &	&	&20	&0.388	&0.966	\\
  \hline
  \hline
\end{tabular}
\end{threeparttable}
\end{center}
\end{table}
\end{landscape}

\begin{landscape}
\begin{table}[]
\caption{Simulation Results for Likelihood Ratio Test under Poisson Regression.}\label{t:LRT2}
\begin{center}
\renewcommand\arraystretch{1}
\begin{threeparttable}[b]
\begin{tabular}{c|cccccccccccc}
\hline
\hline
\multirow{2}{*}{Storing Strategy} &\multirow{2}{*}{$N$} &\multicolumn{5}{c}{Size} & &\multicolumn{5}{c}{Power} \\
\cline{3-13}
& &GO	&OS	&pilot\%	&Pilot	&One-step & &GO	&OS	&pilot\%	&Pilot	&One-step \\
\hline
\multirow{9}{*}{Random}
&\multirow{3}{*}{10000}	&\multirow{3}{*}{0.054}	&\multirow{3}{*}{0.056}	&5	&0.042	&0.054	&	&\multirow{3}{*}{0.454}	&\multirow{3}{*}{0.280}	&5	&0.070	&0.452	\\
&	&	&	&10	&0.052	&0.054	&	&	&	&10	&0.100	&0.454	\\
&	&	&	&20	&0.042	&0.054	&	&	&	&20	&0.126	&0.454	\\
\cline{2-13}
&\multirow{3}{*}{20000}	&\multirow{3}{*}{0.052}	&\multirow{3}{*}{0.054}	&5	&0.054	&0.052	&	&\multirow{3}{*}{0.786}	&\multirow{3}{*}{0.504}	&5	&0.108	&0.784	\\
&	&	&	&10	&0.052	&0.052	&	&	&	&10	&0.154	&0.786	\\
&	&	&	&20	&0.048	&0.052	&	&	&	&20	&0.220	&0.786	\\
\cline{2-13}
&\multirow{3}{*}{50000}	&\multirow{3}{*}{0.048}	&\multirow{3}{*}{0.046}	&5	&0.046	&0.048	&	&\multirow{3}{*}{0.984}	&\multirow{3}{*}{0.886}	&5	&0.138	&0.984	\\
&	&	&	&10	&0.048	&0.048	&	&	&	&10	&0.286	&0.984	\\
&	&	&	&20	&0.052	&0.048	&	&	&	&20	&0.476	&0.984	\\
\hline
\multirow{9}{*}{Nonrandom}
&\multirow{3}{*}{10000}	&\multirow{3}{*}{0.054}	&\multirow{3}{*}{0.056}	&5	&0.046	&0.054	&	&\multirow{3}{*}{0.478}	&\multirow{3}{*}{0.050}	&5	&0.076	&0.470	\\
&	&	&	&10	&0.055	&0.054	&	&	&	&10	&0.082	&0.478	\\
&	&	&	&20	&0.055	&0.054	&	&	&	&20	&0.142	&0.478	\\
\cline{2-13}
&\multirow{3}{*}{20000}	&\multirow{3}{*}{0.052}	&\multirow{3}{*}{0.052}	&5	&0.052	&0.052	&	&\multirow{3}{*}{0.774}	&\multirow{3}{*}{0.062}	&5	&0.086	&0.770	\\
&	&	&	&10	&0.054	&0.052	&	&	&	&10	&0.110	&0.774	\\
&	&	&	&20	&0.048	&0.052	&	&	&	&20	&0.206	&0.774	\\
\cline{2-13}
&\multirow{3}{*}{50000}	&\multirow{3}{*}{0.049}	&\multirow{3}{*}{0.046}	&5	&0.052	&0.049	&	&\multirow{3}{*}{0.982}	&\multirow{3}{*}{0.076}	&5	&0.152	&0.982	\\
&	&	&	&10	&0.050	&0.049	&	&	&	&10	&0.266	&0.982	\\
&	&	&	&20	&0.050	&0.049	&	&	&	&20	&0.432	&0.982	\\
  \hline
  \hline
\end{tabular}
\end{threeparttable}
\end{center}
\end{table}
\end{landscape}

\begin{table}[h]
\caption{The Description of Variables Used in the Airline Dataset.}\label{t:description}
\begin{center}
\renewcommand\arraystretch{1.5}
\begin{threeparttable}[b]
\begin{tabular}{cllcc}
\hline
\hline
	&Variable	&Description &Mean &SD	\\
\hline
Response	&Delayed	&\tabincell{l}{Whether the flight is delayed \\or not, 1 for Yes; 0 for No.} &\multicolumn{2}{c}{Yes: 48.07\%; No: 51.93\%} \\
\hline
\multirow{6}{*}{Covariates}
&DepTime	&Actual departure time	&1349.69	&476.92 \\
&CRSArrTime	&Scheduled arrival time	&1490.84	&493.65 \\
&ElapsedTime	&Actual elapsed time	&119.70	&68.49 \\
&Distance	&\tabincell{l}{Distance between the origin \\and destination (in miles)}	&701.45	&551.29 \\
\hline
\hline
\end{tabular}
\end{threeparttable}
\end{center}
\end{table}

\begin{table}[h]
\caption{The Estimation Results for Logistic Regression under Different Methods.  }\label{t:results}
\begin{center}
\renewcommand\arraystretch{1.5}
\begin{threeparttable}[b]
\begin{tabular}{cccc}
\hline
\hline
	&Global	&One-Shot	&One-Step	\\
\hline
Intercept	&-3.440***	&-2.982***	&-2.813***	\\
DepTime ($\times$100)	&0.096***	&0.066***	&0.069***	\\
CRSArrTime ($\times$100)	&-0.028***	&-0.018***	&-0.019***	\\
ActualElapsedTime	&0.058***	&0.066***	&0.059***	\\
Distance ($\times$100)	&-0.683***	&-0.782***	&-0.699***	\\
\hline
Log-likelihood	&-19083793.2	&-19703781.7	&-19451071.1	\\
\hline
\hline
\end{tabular}
\begin{tablenotes}
     \item[1] *** indicates the p.value smaller than 0.001
   \end{tablenotes}
\end{threeparttable}
\end{center}
\end{table}

\end{document}